\documentclass[doublespacing]{elsart}

\usepackage{amssymb}
\usepackage{epsfig}

\begin{document}

\begin{frontmatter}



\title{A spin field effect transistor for low leakage current}


\author{S. Bandyopadhyay$^a$ and M. Cahay$^b$}
\address[label1]{Department of Electrical and Computer Engineering and 
Department of Physics, 
Virginia Commonwealth 
University, Richmond, VA 23284, USA}
\address[label2]{Department of Electrical and Computer Engineering and Computer 
Science, University of Cincinnati, OH 45221, USA}

\begin{abstract}

In a spin field effect transistor, a magnetic field is inevitably present in the 
channel because of the ferromagnetic source and drain contacts. This field 
causes random unwanted spin precession when carriers interact with non-magnetic 
impurities. The randomized spins lead to a large leakage current when the 
transistor is in the ``off''-state,
resulting in significant standby power dissipation. We can counter this effect 
of the magnetic field by engineering the Dresselhaus spin-orbit interaction in 
the channel with a backgate. For realistic device parameters, a nearly perfect 
cancellation is possible, which should result in a low leakage current.

\end{abstract}

 \end{frontmatter}

\pagebreak

Much of the current interest in spintronic transistors  is motivated by a 
well-known device proposal due to Datta and Das \cite{datta} that has now come 
to be known as a Spin Field Effect Transistor (SPINFET). This device consists  
of a  
one-dimensional 
semiconductor 
channel with half-metallic ferromagnetic source and drain contacts  that are 
magnetized along the channel (Fig. 1). Electrons are 
injected from the source with their spins polarized along the channel's axis. 
The spin  
 is then controllably precessed in the channel with a gate voltage that 
modulates the 
Rashba spin-orbit
interaction \cite{rashba}. At the drain end, the transmission probability of the 
electron depends on the component of its spin vector along the channel.  By 
controlling 
the angle of spin precession in the channel with a gate voltage, 
one can control this component, and hence control 
the 
source-to-drain  current. This realizes the basic ``transistor'' action 
\cite{datta1}.

\begin{figure}
\epsfxsize=2.9in
\epsfysize=4.3in
\centerline{\epsffile{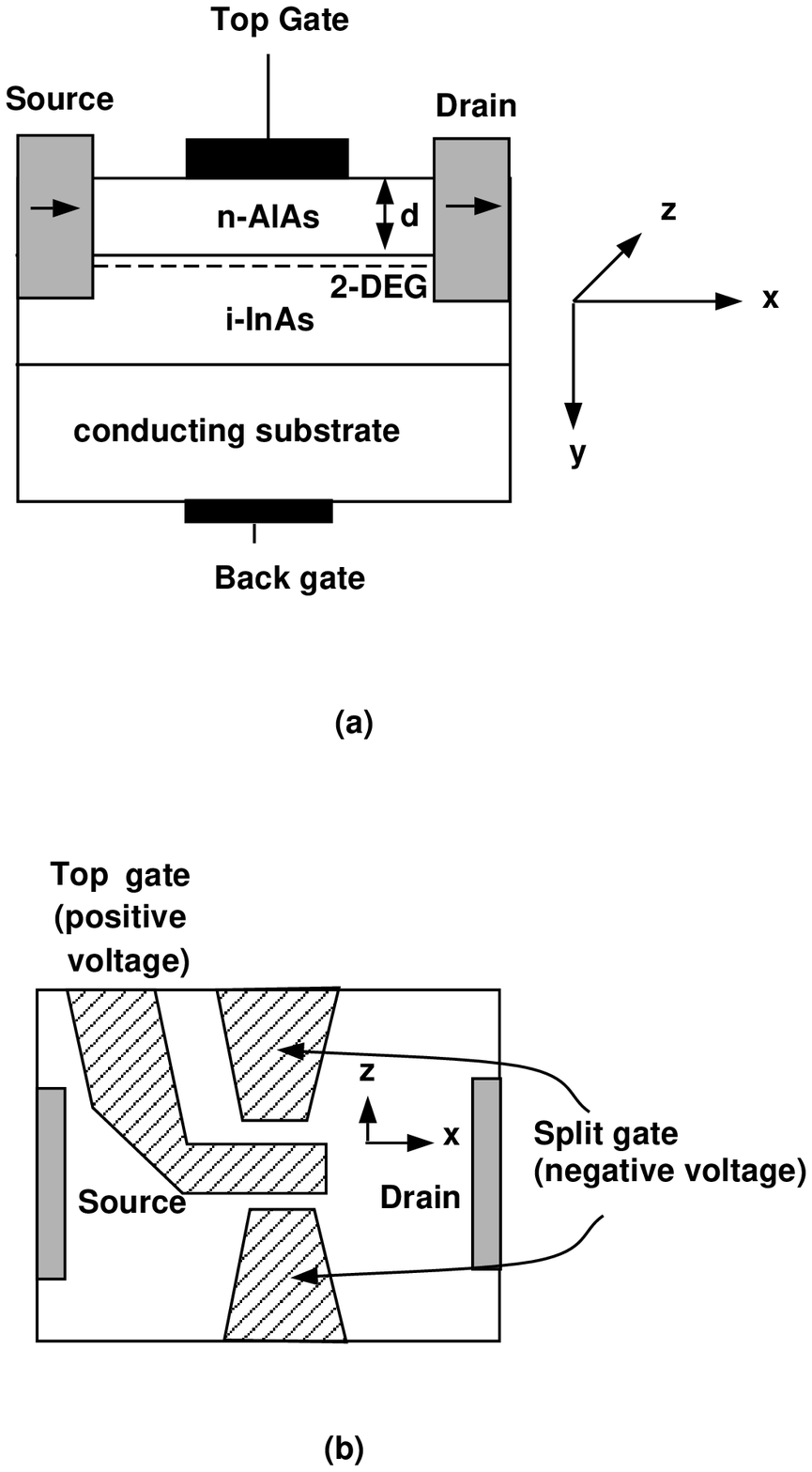}}
\caption[]{\small A spin field effect device with a one-dimensional channel. (a) 
side view showing a top gate for modulating the spin precession via the Rashba 
interaction and a back gate for modulating the channel carrier concentration. 
The substrate will be p$^+$ if we want to deplete the channel with the 
back-gate, and n$^+$ if we want to accumulate it. (b) top view showing the split 
gate configuration required to produce a one-dimensional channel, as well as the 
top gate. A positive voltage is applied to the top gate to increase the 
interface electric field that modulates the Rashba interaction and produces the 
conductance modulation, whereas a negative voltage is applied to the split gates 
to constrict the one-dimensional channel. }
\end{figure}

In their original proposal \cite{datta}, Datta and Das ignored two effects: (i) 
the magnetic field that is inevitably present in the channel because of the 
ferromagnetic source and drain contacts, and (ii) the Dresselhaus spin orbit 
interaction \cite{dresselhaus}  arising from bulk (crystallographic) inversion 
asymmetry.  In the past, we analyzed the effect of the channel magnetic field 
and showed that it could cause weak spin flip scattering via interaction with 
non-magnetic elastic scatterers \cite{cahay_prb}. The flipped spins, whose 
precession angles have been randomized by the spin flip scattering events, will 
lead to a large leakage current when the device is in the ``off''-state. This is 
a serious drawback since it will lead to an unacceptable standby power 
dissipation in a circuit composed of Spin Field Effect Transistors. In order to 
eliminate the leakage current, we must eliminate the unwanted spin flip 
scattering processes. In other words, we must find ways to counter the 
deleterious effect of the channel magnetic field. The purpose of this paper is 
to explore how this can be achieved.

In a strictly one-dimensional structure, where transport in single channeled, 
there is no D'yakonov-Perel spin relaxation \cite{ieee_nano}. Therefore, the 
only agents that can cause spin randomization are hyperfine interactions with 
the nuclei and spin mixing effects caused by the channel magnetic field 
\cite{cahay_prb}. In order to eliminate the latter (which is the stronger of the 
two agents), we can adopt one of two options: either eliminate the magnetic 
field by using non-magnetic spin-injector (source contact) and detector (drain 
contact) \cite{koga}, or counteract the effect of the magnetic field with some 
other effect. The former approach presents a formidable engineering challenge. 
The latter  can be implemented more easily, and, as we show in this paper, is 
achieved by  countering the effect of the magnetic field with the Dresselhaus 
interaction. Calculations based on realistic parameters for InAs transistor 
channels show that this is indeed  possible.

Consider the one-dimensional channel of the device in Fig. 1. Because of the 
magnetized source and drain contacts, a magnetic field exists along the wire in 
the x-direction. We will assume that the channel (x-direction) is along the 
[100] crystallographic axis.

The
effective mass Hamiltonian for the wire, in the Landau gauge 
${\bf A}$ = (0, $-Bz$, 0),  can be written as
\begin{eqnarray}
H & = & (p_x^2 + p_y^2 + p_z^2)/(2m^*) + (e B z p_y)/m^* + (e^2 B^2 z^2)/(2m^*) 
- (g/2) \mu_B B \sigma_x  \nonumber \\
& & + V(y) + V(z) + 2a_{42} [\sigma_x \kappa_x + \sigma_y \kappa_y + \sigma_z 
\kappa_z ] + \eta [ 
(p_x/\hbar) \sigma_z - (p_z/\hbar) \sigma_x ]
\end{eqnarray}
where $g$ is the Land\`e g-factor, $\mu_B$ is the Bohr magneton, $V(y)$ and 
$V(z)$ are the confining potentials along the y- and z-directions, $\sigma$-s 
are the Pauli spin matrices, $2a_{42}$ is the strength of the Dresselhaus 
spin-orbit interaction ($a_{42}$ is a material parameter) and $\eta$ is the 
strength of the 
Rashba spin-orbit 
interaction.

The quantities 
$\kappa$ are defined in ref. \cite{das}.
We will assume that the wire is narrow enough and the temperature is low enough 
that 
only the lowest magneto-electric subband is occupied. Since the Hamiltonian is 
invariant in the $x$-coordinate, the wavevector $k_x$ is a good quantum number 
and the eigenstates are plane waves traveling in the x-direction. Accordingly,  
the spin Hamiltonian (spatial operators are replaced by their expected values)
simplifies to  
\begin{equation}
H = (\hbar^2 k_x^2)/(2 m^*) + E_0 +  (\alpha k_x -  
\beta) 
\sigma_x + \eta k_x \sigma_z
\end{equation}
where $E_0$ is the energy of the lowest magneto-electric subband,   $\alpha(B) 
=  
2a_{42} [ <k_y^2> - <k_z^2>
+ (e^2 B^2 <z^2>/\hbar^2)]$, $\psi(z)$ is the z-component of the wavefunction, 
$\phi(y)$ is the 
y-component of the wavefunction,  $<k_y^2>$ = 
$(1/\hbar^2)<\phi(y)|-(\partial^2/\partial y^2) |\phi(y)>$, $<k_z^2>$ = 
$(1/\hbar^2)<\psi(z)|-(\partial^2/\partial z^2) |\psi(z)>$, and $\beta = (g/2) 
\mu_B B$.

Since the potential $V(z)$ is parabolic ($V(z) = (1/2)m^* \omega_0^2 z^2$), it 
is 
easy to show that $<k_z^2> = m^* \omega/(2 \hbar)$ and $<z^2> = \hbar/( 2 m^* 
\omega)$ where $\omega^2 = \omega_0^2 + \omega_c^2$ and $\omega_c$ is the 
cyclotron frequency ($\omega_c = e B/m^*$). Furthermore, $E_0$ = $(1/2) \hbar 
\omega + E_{\Delta}$ where $E_{\Delta}$ is the energy of 
the lowest subband in the triangular well $V(y)$.

Diagonalizing this Hamiltonian in a truncated Hilbert space spanning the two 
spin resolved states in the lowest subband yields the eigenenergies 
\cite{superlattice}
\begin{equation}
E_{\pm} = {{\hbar^2 k_x^2}\over{2 m^*}}  + E_0 \pm 
\sqrt{\left ( \eta^2 + \alpha^2 \right ) \left ( k_x - {{\alpha 
\beta}\over{\eta^2 + \alpha^2}} \right)^2 + {{\eta^2}\over{\eta^2 + 
\alpha^2}} 
\beta^2} 
\label{eigenenergy}
\end{equation}
and the corresponding eigenstates
\begin{eqnarray}
{\Psi}_{+}(B, x) =
\left [ \begin{array}{c}
             cos(\theta_{k_x})\\
             sin(\theta_{k_x}) \\
             \end{array}   \right ]
             e^{i  k_x x}
             ~~~~~~~~~
{\Psi}_{-}(B, x) =
 \left [ \begin{array}{c}
              sin(\theta_{k_x})\\
               - cos(\theta_{k_x})\\
             \end{array}   \right ]
             e^{i  k_x x}
\label{eigenstate}
\end{eqnarray}
where $\theta_{k_x}$ = $(1/2) arctan [(\alpha k_x - \beta)/\eta k_x]$. 

The dispersion relations given by Equation (\ref{eigenenergy}) can be found 
plotted in ref. \cite{superlattice}. The dispersions are clearly 
nonparabolic and could be {\it asymmetric} about the energy axis. More 
importantly, 
note that the eigenspinors given in Equation 
(\ref{eigenstate}) are functions of $k_x$ because $\theta_{k_x}$ depends on 
$k_x$. Therefore, the eigenspinors are not fixed in any subband, but change with 
$k_x$. In other words, neither subband has a definite spin quantization axis and 
the orientation of the spin vector of an electron in either subband depends 
on the wavevector. Consequently, it is always possible to find two states in the 
two subbands with non-orthogonal spins. Any non-magnetic scatterer (impurity, 
phonon, etc.) can then couple these two states and cause a spin-relaxing 
scattering event. {\it It is this spin flip process that leads to a non-zero 
off-conductance (and leakage current) and needs to be eliminated}.

It is easy to see that the way to eliminate the spin flip process is to enforce 
the condition:
\begin{equation}
\alpha k_x = \beta
\label{condition}
\end{equation}

In this case, the dispersion relations become 
\begin{equation}
E_{\pm} =  E_0 - {{\hbar^2}\over{2 m^*}}k_R^2 + {{\hbar^2}\over{2 m^*}} \left ( 
k_x \pm k_R \right )^2
\label{eigenenergy1}
\end{equation}
where $k_R$ = $m^* \eta/\hbar^2$, and the eigenstates become
\begin{eqnarray}
{\Psi}_{+}( x) =
\left [ \begin{array}{c}
             1 \\
             0 \\
             \end{array}   \right ]
             e^{i  k_x x}
             ~~~~~~~~~
{\Psi}_{-}(x) =
 \left [ \begin{array}{c}
              0\\
              1\\
             \end{array}   \right ]
             e^{i  k_x x}
\label{eigenstate1}
\end{eqnarray}

The dispersion relation in Equation (\ref{eigenenergy1}) is parabolic (two 
parabolas displaced horizontally from the origin 
by $\pm k_R$) and each has a definite (wavevector-independent) spin quantization 
axis which is +z-polarized in the first subband and -z-polarized in the second 
subband. Since the two subbands have orthogonal spin polarizations {\it at any 
wavevector}, no non-magnetic scatter can couple them and cause a spin flip 
event.
Therefore, we can successfully {\it eliminate the unwanted spin flip processes} 
when we enforce the condition in Equation (\ref{condition}).

Equations (\ref{eigenenergy1}) and (\ref{eigenstate1}) are the  dispersions 
and eigenstates employed in ref. \cite{datta}. They are correct only if 
we counteract the channel magnetic field with the Dresselhaus interaction as 
embodied by the condition in Equation (\ref{condition}).

We now proceed to estimate realistic values of $\alpha$ and $\beta$ to see
if the condition in Equation (\ref{condition}) can be realized. This equation 
can be recast as 
\begin{equation}
2 a_{42} \left [ <k_y^2> - {{m^* \omega}\over{2 \hbar}} + {{e^2 B^2}\over{2 m^* 
\hbar \omega}} \right ] k_F = {{g \mu_B B}\over{2}}
\label{condition1}
\end{equation}
where we have assumed that $k_x$ = $k_F$, the Fermi wavevector.

We will assume a 0.2 $\mu$m long channel where the magnetic field can be as 
large as 1 Tesla \cite{cahay_prb1,wrobel}. Table 1 lists the parameters used for 
various quantities used in Equation (\ref{condition1}) (along with the  
citations for the sources when appropriate). Using these parameters, we find 
that in order to counter a magnetic field of 1 Tesla through the Dresselhaus 
interaction, we need $k_F$ = 2.47 $\times$ 10$^9$ m$^{-1}$, which corresponds to 
a linear carrier concentration $n_l$ of 1.54 $\times$ 10$^{9}$ m$^{-1}$. A 
larger magnetic field would require a larger Fermi wavevector and a larger 
carrier concentration.

\bigskip

\begin{table}[h]
\begin{center}
\caption{Parameters for a InAs spin interferometer}
\vskip0.2in
\begin{tabular}{|c|c|}
\hline
$a_{42}$ & 1.6 $\times$ 10$^{-29}$ eV-m$^3$ \cite{jusserand} \\
|g| & 14.4 \cite{bassani} \\
$\hbar \omega_0$ & 10 meV \cite{snyder} \\
$m^*$ & 0.034 $\times$ 9.1 $\times$ 10$^{-31}$ Kg \\
$<k_y^2>$ & 10$^{16}$ m$^{-1}$ \\
\hline
\end{tabular}
\end{center}
\end{table}

\bigskip

The purpose of the backgate in Fig. 1 now becomes clear. We can tune the carrier 
concentration and Fermi wavevector $k_F$ in the channel to the optimum value 
with a backgate voltage. The top gate can then be used exclusively to modulate 
the Rashba interaction which leads to conductance modulation of the transistor. 
As we swing the top gate voltage to switch the transistor from ``on'' to 
``off'', or vice versa, this gate voltage swing $\Delta V_G$ will also induce 
some unavoidable fluctuation in $k_F$. We need to ensure that this fluctuation 
$\Delta k_F$ is a small percentage of $k_F$, so that the act of switching the 
device does not  nullify the balance between the Zeeman splitting (magnetic 
field) and the Dresselhaus interaction.

In ref. \cite{reexam}, we found that $\Delta V_G$ required to induce a spin 
precession of $\pi$ radians in an {\it ideal} 0.2 $\mu$m long InAs SPINFET 
channel is about 50 mV if the gate insulator thickness $d$ (see Fig. 1) is 20 nm 
\cite{theory}. This is the voltage swing required to switch such a SPINFET from 
``on'' to ``off'', or vice versa. Using standard metal oxide semiconductor field 
effect device theory, the change in the 
(two-dimensional) carrier concentration $\Delta N_S$ induced by a gate voltage 
swing $\Delta V_G$  is given by
\begin{equation}
e \Delta N_s = (\epsilon/d) \Delta V_G
\end{equation}
Assuming that the gate insulator is AlAs, for which $\epsilon$ = 8.9 times the 
permittivity of free space \cite{lockwood} and $d$ = 20 nm,
we find that  for $\Delta V_G$ = 50 mV, $\Delta N_s$ = 1.375 $\times$ 10$^{15}$ 
m$^{-2}$. The corresponding fluctuation in the linear carrier concentration is 
found by multiplying this quantity with the effective width $W_{eff}$ of the 
InAs channel which is $\sqrt{\hbar/(2m^* \omega)}$. For $\hbar \omega$ = 10 meV, 
$W_{eff}$  = 22 nm. Therefore the fluctuation in the linear carrier 
concentration   $\Delta n_l$, sustained during switching the SPINFET from ``on'' 
to ``off'', or vice versa, is about 3 $\times$ 10$^7$ m$^{-1}$. This is only  
2\% of $n_l$. Therefore, the modulation of the top gate voltage, during 
switching,  does not affect the channel carrier concentration (or $k_F$) 
significantly. Accordingly, it does not seriously affect the balance between the 
magnetic 
field and the Dresselhaus interaction.

\paragraph{Which way should the contacts be magnetized?} 

Before concluding this paper, we bring out an important issue. It is obvious by 
looking at 
Equations (\ref{eigenenergy}) and (\ref{eigenstate}) that the dispersion
relations and the eigenspinors depend not only on the magnitude but 
also the {\it sign} of $\beta$. Therefore, the ``cancellation effect'' discussed 
in this paper and embodied in Equation (\ref{condition1}) is possible only if 
the magnetization of the contacts is directed in a certain way. For example, if 
the left hand side of Equation (\ref{condition1}) is positive, then the contacts 
should be magnetized along the direction of current flow provided the g-factor 
is positive. If the g-factor is negative, then the contacts should be magnetized 
against the direction of current flow. The exact opposite is true if the left 
hand side of Equation (\ref{condition1}) is negative. To our knowledge, no work 
on the Spin Field Effect Transistor has ever addressed the issue of which way 
the contacts should be magnetized. Here we show, for the first time, that this 
matter is important.

In conclusion, we have shown how the deleterious effect of the channel magnetic 
field in a Spin Field Effect Transistor can be countered with the Dresselhaus 
interaction. In deriving this result, we have also shown that for the 
cancellation to happen, there is an 
optimum channel carrier concentration $n_l$ that depends on the confinement 
energy $\hbar \omega$ in the channel, the channel magnetic field, the strength 
of the Dresselhaus interaction $a_{42}$, the effective mass and Land\'e g-factor 
of the channel material, and the degree of confinement of the two-dimensional 
electron gas at the heterointerface represented by the quantity $<k_y^2>$. Since 
normally many of these parameters will be unknown in any given sample, it will 
be necessary to vary the carrier concentration in the channel with a backgate 
till optimum performance is achieved. To our knowledge, no experimental attempt 
at
demonstrating this device has considered using a backgate to improve 
performance. This may however be an important consideration.

The work of S. B. is supported by the Air Force Office of Scientific Research 
under grant FA9550-04-1-0261 and by the Missile Defense Agency through a 
sub-contract from Wavemat.

\pagebreak

%\bibliographystyle{/d/gady/Style/yzaip}
%\bibliography{/d/rr/latexlib/one}

\begin{thebibliography}{10}
\bibitem{datta1}
Datta and Das appropriately called their device ``an electronic analog of the 
electro-optic modulator'' rather than a ``spin field effect transistor''. This 
device may not have all the attributes of a useful transistor and is probably 
not competitive with state-of-the-art silicon transistors in terms of speed or 
power dissipation (see reference 12).
\end{thebibliography}

\end{document}